\documentclass[conference]{IEEEtran}
\IEEEoverridecommandlockouts
\ifCLASSINFOpdf
  \usepackage[pdftex]{graphicx}
\else
   \usepackage[dvips]{graphicx}
\fi
\usepackage[ruled,vlined]{algorithm2e} 
\usepackage[dvipsnames]{xcolor}
\usepackage{soul,framed}
\usepackage[cmex10]{amsmath}
\usepackage{flushend}
\usepackage{cite}
\usepackage[squaren,Gray]{SIunits}
\usepackage{float}
\usepackage{comment}
\usepackage[noend]{algpseudocode}
\usepackage[utf8]{inputenc}
\usepackage{tikz}
\usepackage{amsmath}
\usepackage{stfloats}
\usepackage{amssymb}
\usepackage{multirow}

\ifCLASSOPTIONcompsoc
    \usepackage[caption=false, font=normalsize, labelfont=sf, textfont=sf]{subfig}
\else
    \usepackage[caption=false, font=footnotesize]{subfig}
\fi

\colorlet{shadecolor}{yellow}

\newcommand{\rank}{L}

\begin{document}

\title{SIM-Aided Near-Field Channel and Localization Estimation With Dimensionality Reduction: A Multiport Network Theory Approach}

\author{
    \IEEEauthorblockN{Andrea~Abrardo,~\IEEEmembership{Senior Member,~IEEE} and Giulio~Bartoli,~\IEEEmembership{Member,~IEEE}}
    
    \IEEEauthorblockA{Department of Information Engineering and Mathematics, University of Siena, Siena, Italy \\
    Email: \{andrea.abrardo, giulio.bartoli\}@unisi.it}
}

\maketitle

\begin{abstract}
The deployment of Extremely Large-Scale Antenna Arrays for 6G enables radiative
near-field sensing but poses significant challenges in terms of hardware
complexity and interference.
Stacked Intelligent Metasurfaces (SIMs) address these limitations by enabling
wave-domain dimensionality reduction.
This paper proposes a rigorous SIM-aided framework for near-field channel and
localization estimation based on Multiport Network Theory, which provides an
electromagnetically consistent characterization accounting for mutual coupling
and non-unilateral inter-layer propagation effects.
An indirect estimation approach is adopted, where the SIM is optimized to perform
analog spatial filtering by projecting the received signal onto a relevant
subspace identified through coarse prior location information.
Within this realistic setting, we analytically characterize the impact of SIM
approximation errors on channel estimation and quantify the resulting effects on
localization performance.
The results show that the proposed architecture preserves the essential wavefront
curvature information required for accurate near-field localization, achieving
performance comparable to fully digital solutions while drastically reducing the
number of radio-frequency chains.
\end{abstract}


\section{Introduction}\label{Intro}
The advent of sixth-generation (6G) mobile networks is pushing operating
frequencies toward the millimeter-wave (mmWave) and Terahertz (THz) bands,
motivating the use of Extremely Large-Scale Antenna Arrays (ELAA) to compensate
for severe path loss.
Such large apertures significantly expand the radiative near-field region
\cite{DardariLIS,LuLargeScaleArray}, where the spherical wavefront curvature
encodes both angular and range information, enabling single-anchor localization
and Integrated Sensing and Communications (ISAC)
\cite{WymeerschRadioLocalization}.
However, ELAA deployments are highly exposed to environmental interference
\cite{SanguinettiElectromagneticInterference}, while fully digital architectures
with one RF chain per antenna element are prohibitively complex and power-hungry.

Reconfigurable Intelligent Surfaces (RIS) have been proposed to shape wireless
channels using passive elements \cite{DiRenzoRIS}, but conventional single-layer
designs offer limited signal processing capabilities.
To overcome these limitations, Stacked Intelligent Metasurfaces (SIM) have been
recently introduced \cite{AnSIM}.
By exploiting multiple diffractive layers, SIMs enable wave-domain processing
and analog signal compression, drastically reducing the number of RF chains with
respect to the number of radiating elements \cite{AnSIMMuBF,AnTransceiverDesign}.

In this work, we propose a SIM-aided framework for joint communications and
sensing in interference-impaired scenarios, where the received signal is
projected onto an equivalent low-dimensional subspace to reduce the number of RF
chains without sacrificing relevant channel information.
By adopting a rigorous multiport network model that captures mutual coupling and
non-unilateral inter-layer propagation \cite{Abrardo_SIM}, and by exploiting
coarse prior channel-related information to identify the relevant signal
subspace, the SIM is configured to implement an adaptive analog projection.
Despite the dimensionality reduction, the physical channel information is
preserved, enabling joint CSI estimation and localization while exploiting the
high spatial resolution of the SIM with significantly lower complexity, cost,
and energy consumption than fully digital architectures.

\subsection{State of the art}
Due to the SIM architecture, the number of digital observation ports $M$ is typically much smaller than the number of SIM elements $K$, in order to achieve substantial reductions in hardware complexity, cost, and power consumption with respect to fully-digital architectures. 
While this design choice is essential to fully exploit the benefits of SIM-based processing, it raises fundamental challenges in Channel State Information (CSI) acquisition, as only a reduced-dimensional projection of the electromagnetic field scattered by the SIM is directly observable in the digital domain. 

Existing works span MMSE-based and subspace-based estimators
\cite{RiceanFading,ChEstWireless}, hybrid digital--wave-domain MSE-minimizing
approaches \cite{MisoChEst}, tensor-based methods for joint channel estimation and
inter-layer calibration \cite{TensorChEst}, as well as Deep Learning techniques
for direct channel inference or refinement of least-squares estimates
\cite{ChEstDL1,ChEstDL2}.

SIMs have also been investigated in ISAC scenarios, from terrestrial networks to
satellite systems \cite{SIMISAC1,SIMISAC2}.
However, SIM-aided localization has received limited attention.
Existing contributions include AoA estimation via 2D-DFT processing
\cite{DiRenzoIFFT}, near-field localization based on Multiport Network Theory
\cite{AbrardoEusipco}, and recent neuromorphic approaches using non-linear SIM
elements \cite{TorcolacciEMProcessing}.

\subsection{Contribution}

A key novelty of this work lies in the adoption of a physically consistent SIM
model based on multiport network theory.
Unlike most existing SIM-related studies, which rely on idealized or abstract
models, the proposed framework explicitly accounts for practical implementation
constraints, including mutual coupling and non-unilateral inter-layer interactions.
Within this realistic setting, we analytically characterize the impact of SIM
approximation errors with respect to the ideal transfer function and derive
theoretical performance bounds that quantify the resulting degradation in channel
estimation accuracy.
Moreover, by leveraging the structure of the near-field channel, which inherently
embeds geometric information through the spherical wavefront, the proposed
approach enables a principled derivation of localization performance bounds
directly from the channel estimate.
Overall, this study provides a novel and realistic assessment of the potential of
SIMs for channel estimation and positioning applications, offering clear insight
into their achievable performance under practical implementation constraints.

\begin{figure}
	\centering
	\includegraphics[trim=0 75 0 85,clip,width=1.0\columnwidth]{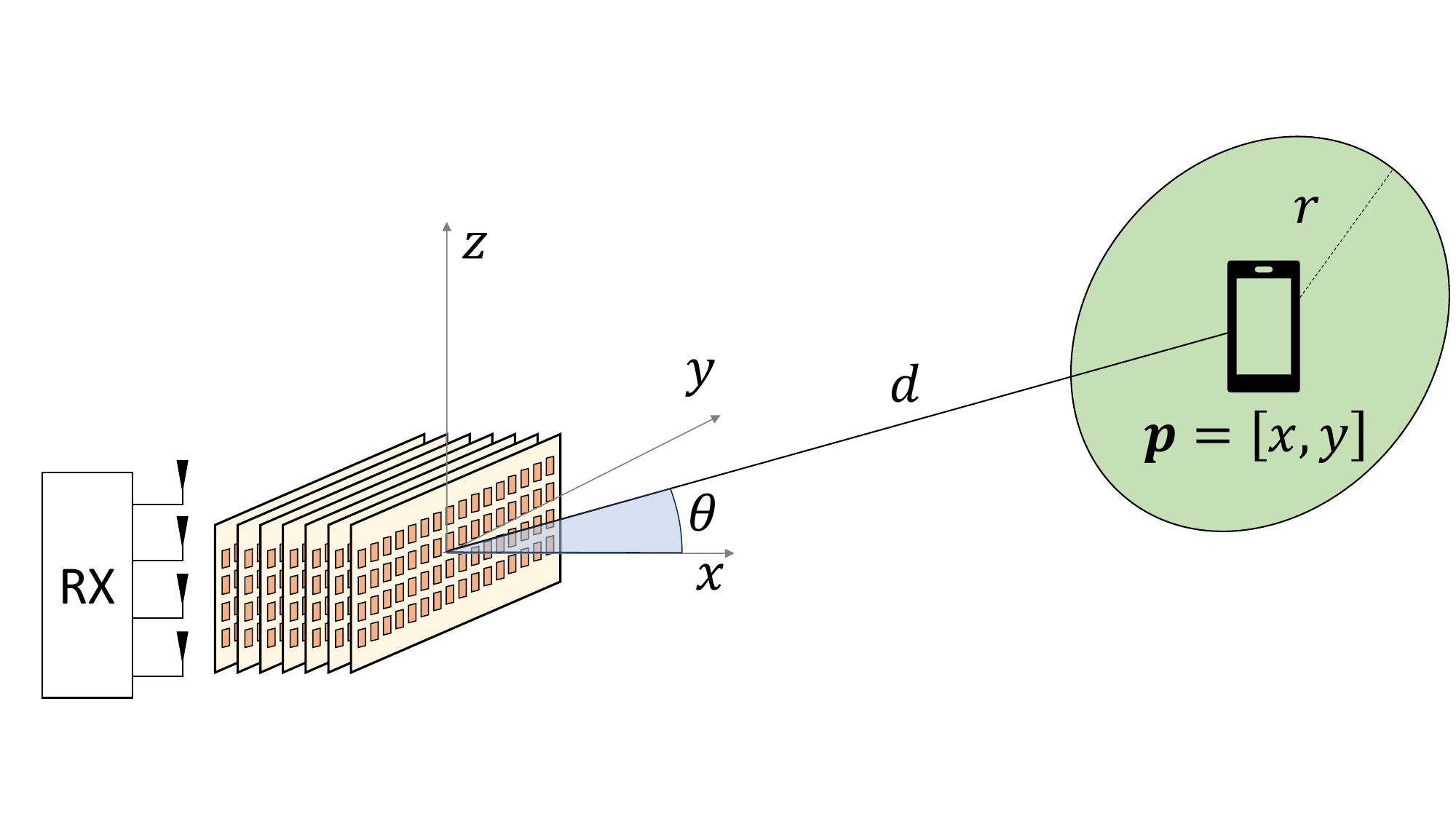}
	\caption{Communication scenario.}
	\label{Fig:Scenario}
\end{figure} 

\section{System Model}\label{sec:SystemModel}
We consider the communication scenario depicted in Fig.~\ref{Fig:Scenario}, involving a SIM-aided receiver and a transmitting node operating in the radiative near-field under environmental electromagnetic interference.

The receiver is equipped with a stacked intelligent metasurface (SIM) composed of $Q$ diffractive layers.
The input layer of the SIM is centered at
$\mathbf{p}_0=(0,0)$ and consists of $K$ scattering elements.
The SIM performs an analog-domain transformation of the impinging electromagnetic field, followed by signal acquisition through $M$ receiver antennas, each connected to an individual RF chain, with $M \ll K$. A single-antenna node communicates with the receiver under line-of-sight (LoS)
conditions.
Its position is $\mathbf{p}=[x,y]$, with $x>0$, and it satisfies
\begin{equation}
  \lVert \mathbf{p}-\mathbf{p}_0\rVert < \frac{2D^2}{\lambda},
\end{equation}
where $D$ denotes the maximum aperture of the SIM input layer and $\lambda$ is the
carrier wavelength.
Under this condition, the link operates in the radiative near-field region, and
spherical wavefront propagation must be explicitly accounted for.

In addition to the signal transmitted by the intended node, the receiver is
assumed to be affected by electromagnetic interference (EMI), encompassing
natural, intentional, and unintentional sources, including background radiation
and residual multipath components not captured by the dominant line-of-sight
(LoS) model.
Due to the short transmitter–receiver distance, thermal noise is neglected and
EMI is considered the dominant impairment.

Although propagation is assumed to be LoS, the received signal amplitude is
modeled as unknown. In near-field localization, position information is primarily conveyed by the
deterministic phase evolution induced by the spherical wavefront, whereas
amplitude variations provide limited and unreliable geometric information.
Accordingly, the channel gain is treated as an unknown nuisance parameter, which
also enables the proposed model to accommodate residual shadowing effects.

The baseband received signal at the first SIM layer is modeled as
\begin{equation}\label{eq:RecSignal}
    \mathbf{r}=\mathbf{h}s+\mathbf{z},
\end{equation}
where $\mathbf{r}\in\mathbb{C}^K$ denotes the received signal, $\mathbf{h}\in\mathbb{C}^K$
is the channel vector, $s\in\mathbb{C}$ is the transmitted symbol, and
$\mathbf{z}\sim\mathcal{CN}(\mathbf{0}_K,\sigma_{\mathbf z}^2\mathbf I)$ represents
isotropic electromagnetic interference. The SIM applies a linear transformation to the received signal, yielding
\begin{equation}\label{eq:LinTrasfSIM}
    \mathbf{y}=\mathbf V\mathbf r,
\end{equation}
where $\mathbf V$ denotes the operator projecting $\mathbf r$ onto a reduced-dimensional
subspace.


\subsection{Transmitting node position and channel model}
Let $\mathbf p_k$ be the Cartesian coordinates of the $k$-th SIM input port.
The distance between the transmitting node at position $\mathbf p$ and the $k$-th
port is
\begin{equation}
  d_k(\mathbf p)=\|\mathbf p-\mathbf p_k\|_2,\qquad k=1,\ldots,K.
\end{equation}
Assuming a LoS near-field channel with deterministic phase evolution and unknown
gain and phase offsets, the channel coefficient at the $k$-th element is modeled as
\begin{equation}\label{eq:ch_model}
  h_k(\mathbf p,G,\theta)
  =G\,e^{j\theta}\exp\!\left(-j\frac{2\pi}{\lambda}d_k(\mathbf p)\right),
\end{equation}
where $\lambda$ is the wavelength, $G\in\mathbb{R}^+$ captures shadowing and path-loss
uncertainty, and $\theta\in[0,2\pi)$ accounts for phase synchronization offsets.
Stacking all coefficients yields
\begin{equation}
  \mathbf h(\mathbf p,G,\theta)=G e^{j\theta}\mathbf a(\mathbf p),
\end{equation}
with 
$\mathbf a(\mathbf p)=\big[e^{-j\frac{2\pi}{\lambda}d_1(\mathbf p)},\ldots,
e^{-j\frac{2\pi}{\lambda}d_K(\mathbf p)}\big]^{\mathsf T}$ is the near-field steering vector.

We assume that the receiver has prior knowledge of a spatial uncertainty region
$\mathcal P$ containing the transmitter, described by the PDF
$f_{\boldsymbol{\wp}}(\mathbf p)$.
When $\mathcal P$ spans multiple wavelengths, the rapid phase variations in
$\mathbf h$ allow modeling the channel as a zero-mean proper complex
Gaussian random vector \cite{Abrardo_two}, i.e.,
\[
\mathbf h\sim\mathcal{CN}(\mathbf 0,\mathbf{R}_\mathbf{h}).
\]
The channel covariance matrix is obtained by averaging over the spatial and gain
distributions as
\begin{equation}\label{eq:ch_stats1}
\mathbf{R}_\mathbf{h}
=\mathbb E_{G,\mathbf p}[\mathbf h\mathbf h^H]
\approx
\sigma_G^2
\int_{\mathcal P}
\mathbf a(\mathbf p)\mathbf a^H(\mathbf p)
f_{\boldsymbol{\wp}}(\mathbf p)\,d\mathbf p,
\end{equation}
where $\sigma_G^2=\mathbb E[G^2]$.
Knowledge of $f_{\boldsymbol{\wp}}(\mathbf p)$ thus directly enables the computation
of $\mathbf{R}_\mathbf{h}$, which underpins the proposed SIM optimization framework.


\subsection{SIM Multiport Network Model and optimization approach}
We adopt the multiport network model introduced in \cite{Abrardo_SIM} to
characterize the Stacked Intelligent Metasurface (SIM).
As shown in \cite{Abrardo_SIM}, this modeling framework enables an
electromagnetically consistent description of SIM architectures, explicitly
capturing mutual coupling among elements and non-unilateral inter-layer propagation effects,
thereby overcoming the strong unilateral propagation assumptions commonly
adopted in the vast majority of existing SIM-related works.

The SIM is modeled as a layered structure composed of $Q$ transmitting RIS
(T-RIS) layers, each consisting of $K$ unit cells.
Mathematically, it is represented as a $2KQ$-port network characterized by a
static impedance matrix $\mathbf{Z}_{SS}\in\mathbb{C}^{2QK\times 2QK}$, which
accounts for structural electromagnetic interactions and mutual coupling, and by
a tunable diagonal impedance matrix $\mathbf{Z}_{S}(\boldsymbol{\eta})$.
Specifically, the overall electromagnetic response of the SIM is governed by the phase shifts
applied at the ports of the transmitting RIS elements, collected in the vector
$\boldsymbol{\eta}$ of $KQ$ tunable parameters.

The resulting input–output relationship is given by \cite{Abrardo_SIM}
\begin{equation}\label{eq:SIM_IO_0}
  \mathbf{y}
  =
  \mathbf{A}\mathbf{T}(\boldsymbol{\eta})\mathbf{r},
  \qquad
  \mathbf{T}(\boldsymbol{\eta})
  =
  \bigl(\mathbf{Z}_{SS}+\mathbf{Z}_{S}(\boldsymbol{\eta})\bigr)^{-1},
\end{equation}
where $\mathbf{r}$ denotes the incident signal, $\mathbf{y}$ the signal observed
at the receiver ports, and $\mathbf{A}$ models the coupling between the last SIM
layer and the receiver antennas.
In the considered framework, the effective SIM transfer function satisfies
\[
\mathbf{A}\mathbf{T}(\boldsymbol{\eta}) \equiv \mathbf{V},
\]
i.e., the SIM is configured to approximate the desired linear transformation
$\mathbf{V}$ introduced in the previous section.

The SIM configuration $\boldsymbol{\eta}$ is obtained by minimizing the mismatch
between the actual SIM response and a desired target mapping, e.g., a prescribed
subspace projection.
Thanks to the differentiable structure of the multiport model, the associated
optimization problem can be efficiently solved using reduced-complexity
gradient-based algorithms akin to back-propagation, despite the fully
electromagnetic nature of the model, as detailed in \cite{Abrardo_SIM}.
This approach ensures that the resulting SIM configuration is both physically
realizable and consistent with the underlying electromagnetic constraints.

\section{Channel and Position Estimation}

\subsection{Channel estimation}
Assuming, without loss of generality, that the transmitter sends the symbol $s=1$, the received signal model can be written as
\begin{equation}\label{eq:full_model}
\mathbf r = \mathbf h + \mathbf z.
\end{equation}

Let the eigendecomposition of the channel covariance matrix be
\begin{equation}\label{eq:Rh_eig}
\mathbf{R}_\mathbf{h} = \mathbf U_t \mathbf D_t \mathbf U_t^H,
\end{equation}
where $\mathbf D_t$ is diagonal and $\mathbf U_t$ is unitary. Since $\mathbf h$ and $\mathbf r$ are jointly Gaussian, the MMSE estimator of $\mathbf h$ coincides with the LMMSE estimator and is given by
\begin{equation}\label{eq:mmse_closed_form}
    \widehat{\mathbf h}_{\mathrm{MMSE}} = \mathbf{R}_\mathbf{h} \bigl(\mathbf{R}_\mathbf{h} + \sigma_z^2 \mathbf I_K\bigr)^{-1} \mathbf r.
\end{equation}

Substituting \eqref{eq:Rh_eig} into \eqref{eq:mmse_closed_form} yields the spectral-domain representation, i.e.,
\begin{equation}\label{eq:mmse_spectral}
    \widehat{\mathbf h}_{\mathrm{MMSE}} = \mathbf U_t \mathbf D_t \bigl(\mathbf D_t + \sigma_z^2 \mathbf I_K\bigr)^{-1} \mathbf U_t^H \mathbf r.
\end{equation}
When the uncertainty region $\mathcal{P}$ is spatially limited, the channel covariance matrix $\mathbf{R}_\mathbf{h}$ is typically rank-deficient. Let
\[
\rank = \mathrm{rank}(\mathbf{R}_\mathbf{h}) < K,
\]
and let $\mathbf D \in \mathbb{R}_+^{\rank \times \rank}$ collect the strictly positive eigenvalues of $\mathbf{R}_\mathbf{h}$, with corresponding eigenvectors forming the matrix $\mathbf U \in \mathbb{C}^{K \times \rank}$.
It follows that $\mathbf U^H \mathbf U = \mathbf I_{\rank}$, while $\mathbf U \mathbf U^H \neq \mathbf I_K$ for {$\rank < K$}.
Then, \eqref{eq:mmse_spectral} becomes
\begin{equation}\label{eq:mmse_spectral2}
    \widehat{\mathbf h}_{\mathrm{MMSE}} = \mathbf U \mathbf D \bigl(\mathbf D + \sigma_z^2 \mathbf I_{\rank}\bigr)^{-1} \mathbf U^H \mathbf r.
\end{equation}
This expression shows that MMSE estimation can be equivalently performed on the sufficient statistic
\begin{equation}\label{eq:reduced_stat}
\mathbf y \triangleq \mathbf U^H \mathbf r
= \mathbf U^H \mathbf h + \widetilde{\mathbf z},
\end{equation}
where $\widetilde{\mathbf z} \triangleq \mathbf U^H \mathbf z \sim \mathcal{CN}(\mathbf 0, \sigma_z^2 \mathbf I_{\rank})$.

An alternative estimator that does not require knowledge of the noise variance $\sigma_z^2$ is the reduced-subspace least-squares (RS-LS) estimator, defined as
\begin{equation}\label{eq:RS-LS est}
\widehat{\mathbf h}_{\mathrm{RS\text{-}LS}}
=
\mathbf U \mathbf y.
\end{equation}

\subsection{Position estimation bounds}
\label{sec:position-estimation}

After channel estimation, the resulting observation model is
\begin{equation}
  \hat{\mathbf h} = \mathbf h(\boldsymbol{\epsilon}) + \mathbf n,
\end{equation}
where $\mathbf n\sim\mathcal{CN}(\mathbf 0,\sigma_n^2\mathbf I_K)$ represents the
residual estimation noise and $\boldsymbol{\epsilon}$ is the unknown real-valued parameter vector
\begin{equation}
  \boldsymbol{\epsilon} =
  \begin{bmatrix}
    x & y & G & \theta
  \end{bmatrix}^{\mathsf T},
\end{equation}
where $(x,y)$ denote the transmitter position, while $(G,\theta)$ represent
nuisance parameters associated with the channel gain and phase.
Assuming that the position is recovered indirectly from the channel estimate,
a bound on the achievable positioning accuracy is provided by the classical
(non-Bayesian) Position Error Bound (PEB), derived conditionally on the resulting
observation model.

To elaborate, under the assumed Gaussian model, the Fisher Information Matrix (FIM) associated
with $\boldsymbol{\epsilon}$ is given by
\begin{equation}
  \mathbf I(\boldsymbol{\epsilon})
  =
  \frac{1}{\sigma_n^2}
  \Re\!\left\{
    \mathbf J(\boldsymbol{\epsilon})^H
    \mathbf J(\boldsymbol{\epsilon})
  \right\},
\end{equation}
where $\mathbf J(\boldsymbol{\epsilon})$ denotes the Jacobian of the channel with
respect to the unknown parameters. The entries of $\mathbf J(\boldsymbol{\epsilon})$ can be straightforwardly
derived in closed form from the channel model in \eqref{eq:ch_model}.
The covariance matrix of any unbiased estimator of $\boldsymbol{\epsilon}$ is
lower bounded by $\mathbf I(\boldsymbol{\epsilon})^{-1}$. Since only the spatial
coordinates $(x,y)$ are of interest for localization, the corresponding Position
Error Bound (PEB) is obtained by extracting the submatrix of
$\mathbf I(\boldsymbol{\epsilon})^{-1}$ associated with the position parameters.

\section{SIM-Based Reduced Space Channel Estimation}
\subsection{Channel estimation}
As shown in the previous section, the optimal information-preserving linear
transformation for electromagnetic-domain dimensionality reduction is
\begin{equation}\label{eq:V_opt}
  \mathbf V = \mathbf U^H .
\end{equation}

In practice, however, the linear transformation implemented by the SIM is
constrained by the physical structure of the metasurface and can only
approximate the ideal digital projection in \eqref{eq:V_opt}.
Assume that the SIM implements a linear transformation
\begin{equation}\label{eq:Err_SIM}
  \mathbf V = \mathbf U^H + \boldsymbol{\Delta},
\end{equation}
where $\boldsymbol{\Delta}\in\mathbb C^{L\times K}$ represents a deterministic mismatch
with respect to the ideal projection $\mathbf U^H$. Under the SIM observation model in \eqref{eq:Err_SIM} the channel estimators in \eqref{eq:mmse_spectral2} and \eqref{eq:RS-LS est},
which relied on the ideal projection $\mathbf V=\mathbf U^H$, must be revised.

To elaborate, considering $\mathbf h\sim\mathcal{CN}(\mathbf 0,\mathbf{R}_\mathbf{h})$, the MMSE estimate of $\mathbf h$
from $\mathbf y = \mathbf{V}\mathbf{r}$ is
\begin{equation}\label{eq:mmse_after_sim}
  \widehat{\mathbf h}_{\mathrm{MMSE}}
  =
  \mathbf{R}_\mathbf{h} \mathbf V^H
  \Bigl(
    \mathbf V\mathbf{R}_\mathbf{h}\mathbf V^H + \sigma_z^2\,\mathbf V\mathbf V^H
  \Bigr)^{-1}
  \mathbf y .
\end{equation}
The corresponding MMSE error covariance matrix is the standard posterior
covariance
\begin{equation}\label{eq:mmse_errcov_after_sim}
  \mathbf C_{\mathrm{MMSE}}
  =
  \mathbf{R}_\mathbf{h}
  -
  \mathbf{R}_\mathbf{h}\mathbf V^H
  \Bigl(
    \mathbf V\mathbf{R}_\mathbf{h}\mathbf V^H + \sigma_z^2\,\mathbf V\mathbf V^H
  \Bigr)^{-1}
  \mathbf V\mathbf{R}_\mathbf{h},
\end{equation}
and the associated scalar MSE is
\begin{equation}\label{eq:mmse_mse_after_sim}
  \mathrm{MSE}_{\mathrm{MMSE}}
  =
  \mathbb{E}\!\left[\|\mathbf h-\widehat{\mathbf h}_{\mathrm{MMSE}}\|_2^2\right]
  =
  \mathrm{tr}\!\left(\mathbf C_{\mathrm{MMSE}}\right).
\end{equation}
We now consider the reduced-subspace least-squares (RS-LS) channel estimator
operating after the SIM.
Assuming that the channel lies in the reduced subspace, i.e.,
$\mathbf h=\mathbf U\mathbf g$, and based on the post-SIM observation model, the RS-LS estimate of the channel is given by
\begin{equation}\label{eq:rslS_after_sim}
  \widehat{\mathbf h}_{\mathrm{RS\text{-}LS}}
  =
  \mathbf U\,\widehat{\mathbf g}_{\mathrm{LS}},
  \qquad
  \widehat{\mathbf g}_{\mathrm{LS}}
  =
  \bigl(\mathbf U^H\mathbf V^H\mathbf V\mathbf U\bigr)^{-1}
  \mathbf U^H\mathbf V^H \mathbf y .
\end{equation}
Let $\mathbf A \triangleq \mathbf V\mathbf U \in \mathbb{C}^{L\times r}$, where
$r=\mathrm{rank}(\mathbf{R}_\mathbf{h})$.
Then $\mathbf y=\mathbf A\mathbf g+\mathbf V\mathbf z$, and the noise covariance is
$\mathbf C_z=\sigma_z^2\,\mathbf V\mathbf V^H$.
The error covariance of $\widehat{\mathbf g}_{\mathrm{LS}}$ is
\begin{equation}\label{eq:rslS_g_errcov}
  \mathbf C_{\mathrm{LS},g}
   =
  (\mathbf A^{H}\mathbf A)^{-1}\mathbf A^{H}\mathbf C_z\,\mathbf A
  (\mathbf A^{H}\mathbf A)^{-1},
\end{equation}
and the corresponding channel-domain error covariance is
\begin{equation}\label{eq:rslS_h_errcov}
  \mathbf C_{\mathrm{RS\text{-}LS}}
  =
  \mathbb{E}\!\left[
    (\mathbf h-\widehat{\mathbf h}_{\mathrm{RS\text{-}LS}})
    (\mathbf h-\widehat{\mathbf h}_{\mathrm{RS\text{-}LS}})^{H}
  \right]
  =
  \mathbf U\,\mathbf C_{\mathrm{LS},g}\,\mathbf U^{H}.
\end{equation}
The associated scalar MSE is
\begin{equation}\label{eq:rslS_mse_after_sim}
  \mathrm{MSE}_{\mathrm{RS\text{-}LS}}
  =
  \mathbb{E}\!\left[\|\mathbf h-\widehat{\mathbf h}_{\mathrm{RS\text{-}LS}}\|_2^2\right]
  =
  \mathrm{tr}\!\left(\mathbf C_{\mathrm{RS\text{-}LS}}\right)
  =
  \mathrm{tr}\!\left(\mathbf C_{\mathrm{LS},g}\right),
\end{equation}
where the last equality follows from $\mathbf U^{H}\mathbf U=\mathbf I_r$.

\subsection{Sensitivity to SIM approximation error}

We quantify the approximation error through two complementary metrics.
The first one is the \emph{relative Frobenius mismatch}
\begin{equation}
  \delta_{\mathrm{rel}}
  \triangleq
  \frac{\|\mathbf V-\mathbf U^H\|_F}{\|\mathbf U^H\|_F}
  =
  \frac{\|\boldsymbol{\Delta}\|_F}{\sqrt{L}},
\end{equation}
which provides a normalized energy measure of the SIM approximation error.
The second one is the \emph{effective subspace mismatch}
\begin{equation}
  \delta_U \triangleq \|\boldsymbol{\Delta}\mathbf U\|_2.
\end{equation}

Given the reduced observation operator $\mathbf A \triangleq \mathbf V\mathbf U \in \mathbb C^{L\times L}$ and the associated matrix
\begin{equation}
  \mathbf G(\mathbf V)
  \triangleq
  \mathbf A^H\mathbf A
  =
  \mathbf U^H\mathbf V^H\mathbf V\mathbf U,
\end{equation}
substituting $\mathbf V=\mathbf U^H+\boldsymbol{\Delta}$ yields
\begin{equation}
  \mathbf A
  =
  (\mathbf U^H+\boldsymbol{\Delta})\mathbf U
  =
  \mathbf I_L + \boldsymbol{\Delta}\mathbf U,
\end{equation}
and therefore
\begin{equation}\label{eq:G_perturb_correct}
  \mathbf G(\mathbf V)
  =
  \mathbf I_L
  + \boldsymbol{\Delta}\mathbf U
  + \mathbf U^H\boldsymbol{\Delta}^H
  + \mathbf U^H\boldsymbol{\Delta}^H\boldsymbol{\Delta}\mathbf U
  \;\triangleq\;
  \mathbf I_L + \mathbf E .
\end{equation}

Using the triangle inequality we have
\begin{equation}\label{eq:E_bound_final}
  \|\mathbf E\|_2
  \le
  \|\boldsymbol{\Delta}\mathbf U\|_2
  + \|\mathbf U^H\boldsymbol{\Delta}^H\|_2
  + \|\mathbf U^H\boldsymbol{\Delta}^H\boldsymbol{\Delta}\mathbf U\|_2
  \le
  2\delta_U + \delta_U^2 .
\end{equation}
Since $\mathbf G(\mathbf V)=\mathbf I_L+\mathbf E$ and $\mathbf E$ is Hermitian, its
spectral norm coincides with the maximum absolute eigenvalue,
\begin{equation}
  \|\mathbf E\|_2 = \max_i |\lambda_i(\mathbf E)|.
\end{equation}
Therefore, all eigenvalues of $\mathbf G(\mathbf V)$ satisfy
\begin{equation}
  1-(2\delta_U+\delta_U^2)
  \le
  \lambda_i\!\big(\mathbf G(\mathbf V)\big)
  \le
  1+(2\delta_U+\delta_U^2),
  \qquad i=1,\ldots,L.
\end{equation}
In particular, if $2\delta_U+\delta_U^2<1$, then $\mathbf G(\mathbf V)$ is positive
definite.

\paragraph{MSE degradation bound}

Assuming that the SIM preserves energy, i.e., $\mathbf V\mathbf V^H=\mathbf I_L$,
the RS-LS mean-square error is
\begin{equation}
  \mathrm{MSE}_{\mathrm{RS\text{-}LS}}(\mathbf V)
  =
  \sigma_z^2\,\mathrm{tr}\!\left(\mathbf G(\mathbf V)^{-1}\right).
\end{equation}
In the ideal case $\mathbf V=\mathbf U^H$, one has
$\mathbf G(\mathbf V)=\mathbf I_L$ and
\begin{equation}
  \mathrm{MSE}_0 = \sigma_z^2 L .
\end{equation}
Since the trace of $\mathbf G^{-1}$ is the sum of the reciprocals of the eigenvalues
of $\mathbf G$, and each reciprocal is upper bounded by the inverse of the minimum
eigenvalue, the relative MSE degradation satisfies
\begin{equation}\label{eq:MSE_ratio_final}
  \frac{\mathrm{MSE}_{\mathrm{RS\text{-}LS}}(\mathbf V)}{\mathrm{MSE}_0}
  \le
  \frac{1}{1-(2\delta_U+\delta_U^2)} .
\end{equation}

\section{Numerical Results}
Due to the two-dimensional geometry of the considered scenario shown in Fig.~\ref{Fig:Scenario}, the first layer of the SIM is implemented as a uniform planar array (UPA) with $K_y=64$ elements aligned along the $y$ axis, in order to ensure sufficient phase resolution, and only $K_z=4$ elements along the $z$ axis, whose role is to increase the overall SIM gain.
We consider $Q=7$ layers stacked on the $x$ axis, and all SIM layers are assumed to share the same geometry, resulting in a SIM with a total of $K_yK_zQ=1792$ tunable elements, i.e., a
control vector $\boldsymbol{\eta}$ of dimension $1792$.
The detailed SIM architecture follows the design in \cite{Abrardo_SIM}, where
each SIM element is implemented as a metallic dipole and the carrier frequency
is set $f_0=28~\giga\hertz$; further implementation details are referred to \cite{Abrardo_SIM}.
The SIM configuration is obtained by minimizing the error between the target transformation $\mathbf U^H$ and the electromagnetic response of the SIM.

The receiver is assumed to be composed of identical dipoles, similar to those forming the SIM, arranged in a uniform linear array (ULA) of $L$ elements along the $y$ axis, and placed at a distance of $\lambda$ from the $Q$-th SIM layer.

Under these conditions, the aperture $D$ of the receiving array is approximately $32~\centi\meter$, which corresponds to a Fraunhofer distance of about $20~\meter$.
Simulation results are obtained assuming a log-normal shadowing with a $3~\deci\bel$ standard deviation for the channel gain $G$, and the presence of interference modeled through a noise variance $\sigma_z^2$.
The latter is set according to a prescribed signal-to-noise ratio (SNR) and is defined with respect to the average received signal energy under the considered configurations.

Finally, the results reported in the following are obtained by considering a fixed number of outputs $L=6$, corresponding to a reduction in the number of RF chains from $64$ to $6$ with respect to a fully digital MIMO architecture, and a circular uncertainty region with a diameter of $0.6~\meter$.
In the considered scenarios, the choice $M=6$ is consistent with the effective rank of the channel covariance matrix; however, in some cases, as discussed later, it may lead to an under-dimensioned representation, introducing a performance loss due to channel subspace truncation.
Regarding the SIM optimization, the design procedure is constrained to achieve an effective subspace mismatch not exceeding $\delta_U \le 0.1$.

\begin{figure*}
	\centering
	\subfloat[][$\theta = 0$]{\includegraphics[]{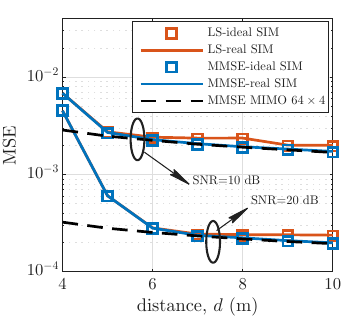}\label{SubFig:mse_vs_d_theta_0000pi}}
	\subfloat[][$\theta = \frac{\pi}{6}$]{\includegraphics[]{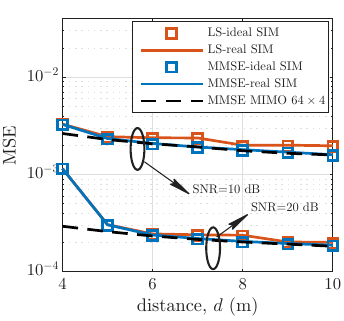}\label{SubFig:mse_vs_d_theta_0167pi}}
	\subfloat[][$\theta = \frac{\pi}{3}$]{\includegraphics[]{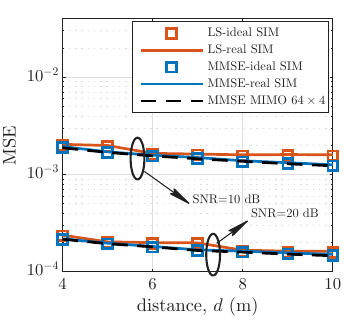}\label{SubFig:mse_vs_d_theta_0333pi}}
		\caption{MSE as a function of the distance for different angles.}
	\label{MultiFig:mse_vs_d}
\end{figure*}

Fig.~\ref{MultiFig:mse_vs_d} shows the MSE of the MMSE and LS channel estimators as a function of the transmitter--receiver distance, considering both ideal and practical SIM implementations.
The ideal case corresponds to a scenario in which the received signal is projected exactly through the transformation $\mathbf U^H$, whereas the practical case refers to the use of a SIM whose electromagnetic response approximates $\mathbf U^H$ with an effective subspace mismatch not exceeding $\delta_U$.
It is worth noting that, in both cases, the projection $\mathbf U^H$ is constructed using the $L=6$ dominant eigenmodes of the channel covariance matrix, so that a residual approximation error is inherently present due to the limited number of outputs.

Each subfigure corresponds to a different angle of arrival, namely $\theta = k\pi/6$ for $k \in \{0,1,2\}$.
Results are reported for two different SNR levels.
Remarkably, the curves corresponding to the ideal and practical SIM implementations are almost indistinguishable, as expected.
It can also be observed that the RS-LS approach achieves slightly worse performance than the MMSE estimator.

For comparison, the performance of a conventional fully digital MIMO receiver with a $64\times 4$ antenna array and an equivalent number of RF chains is also included.
The fully digital MIMO receiver, which implements an MMSE channel estimator, achieves superior performance only at very short distances, where the effective degrees of freedom of the channel exceed $L=6$, whereas its performance converges to that of the SIM-based MMSE solution at larger distances.

\begin{figure*}
	\centering
	\subfloat[][$\theta = 0$]{\includegraphics[]{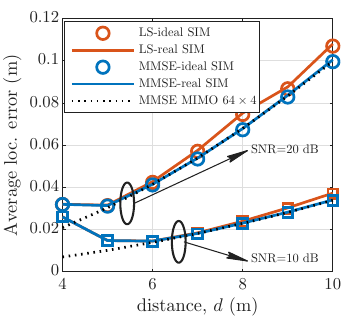}\label{SubFig:loc_vs_d_theta_0000pi}}
	\subfloat[][$\theta = \frac{\pi}{6}$]{\includegraphics[]{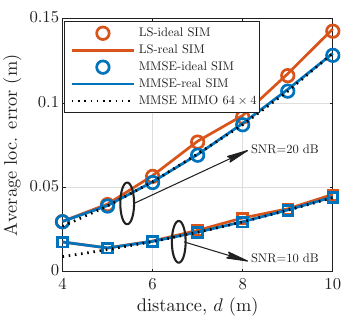}\label{SubFig:loc_vs_d_theta_0167pi}}
	\subfloat[][$\theta = \frac{\pi}{3}$]{\includegraphics[]{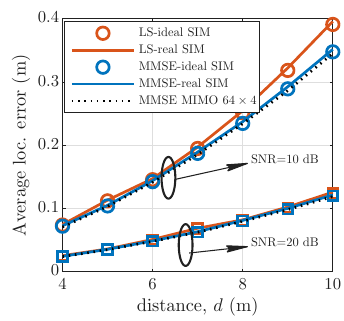}\label{SubFig:loc_vs_d_theta_0333pi}}
		\caption{Location error as a function of the distance for different angles.}
	\label{MultiFig:loc_vs_d}
\end{figure*}

Fig.~\ref{MultiFig:loc_vs_d} reports the localization error following the same structure as Fig.~\ref{MultiFig:mse_vs_d}.
The conclusions are consistent with those drawn from the channel estimation results.
Overall, these results clearly highlight the benefits of SIM-based architectures, which enable large apertures and accurate near-field localization while relying on a limited number of RF chains.

\section{Conclusion}\label{sec:Conclusion}
This paper presented a SIM-aided framework for near-field localization based on Multiport Network Theory.
By optimizing the SIM to perform analog dimensionality reduction, we derived theoretical performance bounds that explicitly characterize the impact of SIM approximation errors on channel estimation.
The resulting analysis shows that the essential wavefront curvature information required for accurate near-field positioning is preserved despite the reduced dimensionality.
Specifically, the derived Position Error Bound (PEB) demonstrates that high-precision localization can be achieved with a minimal number of RF chains, even in the presence of hardware-induced approximations.
Overall, these results validate SIM-based architectures as a scalable and energy-efficient enabler for next-generation 6G integrated sensing and communications systems.

\bibliographystyle{IEEEtran}
\bibliography{biblio}

@ARTICLE{LuLargeScaleArray,
    author = {Lu, Haiquan and Zeng, Yong},
    journal = {{IEEE} Trans. Wireless Commun.}, 
    title = {{Communicating With Extremely Large-Scale Array/Surface: Unified Modeling and Performance Analysis}}, 
    year = {2022},
    volume = {21},
    number = {6},
    pages = {4039--4053}
}

@ARTICLE{WymeerschRadioLocalization,
    author = {Wymeersch, Henk and He, Jiguang and Denis, Benoit and Clemente, Antonio and Juntti, Markku},
    journal = {{IEEE} Veh. Technol. Mag.}, 
    title = {{Radio Localization and Mapping With Reconfigurable Intelligent Surfaces: Challenges, Opportunities, and Research Directions}}, 
    year = {2020},
    volume = {15},
    number = {4},
    pages = {52--61}
}

@ARTICLE{AnSIM,
    author = {An, Jiancheng and Xu, Chao and Ng, Derrick Wing Kwan and Alexandropoulos, George C. and Huang, Chongwen and Yuen, Chau and Hanzo, Lajos},
    journal = {{IEEE} J. Sel. Areas Commun.}, 
    title = {Stacked Intelligent Metasurfaces for Efficient Holographic MIMO Communications in 6G}, 
    year = {2023},
    volume = {41},
    number = {8},
    pages = {2380--2396},
}

@article{Abrardo_SIM,
  author       = {Andrea Abrardo and
                  Giulio Bartoli and
                  Alberto Toccafondi},
  title        = {A Novel Comprehensive Multiport Network Model for Stacked Intelligent
                  Metasurfaces {(SIM)} Characterization and Optimization},
  journal      = {{IEEE} Trans. Commun.},
  volume       = {73},
  number       = {11},
  pages        = {11559--11573},
  year         = {2025},
}

@ARTICLE{Abrardo_two,
	author = {Jiang, Fan and Abrardo, Andrea and Keykhosravi, Kamran and Wymeersch, Henk and Dardari, Davide and Di Renzo, Marco},
	journal = {{IEEE} Trans. Wireless Commun.}, 
	title = {{Two-Timescale Transmission Design and RIS Optimization for Integrated Localization and Communications}}, 
	year = {2023},
	volume = {22},
	number = {12},
	pages = {8587--8602},
}

@ARTICLE{DardariLIS,
  author={Dardari, Davide},
  journal={{IEEE} J. Sel. Areas Commun.}, 
  title={{Communicating With Large Intelligent Surfaces: Fundamental Limits and Models}}, 
  year={2020},
  volume={38},
  number={11},
  pages={2526--2537},
}

@ARTICLE{SanguinettiElectromagneticInterference,
  author={Long, Wen-Xuan and Moretti, Marco and Abrardo, Andrea and Sanguinetti, Luca and Chen, Rui},
  journal={{IEEE} Trans. Wireless Commun.}, 
  title={{MMSE Design of RIS-Aided Communications With Spatially-Correlated Channels and Electromagnetic Interference}}, 
  year={2024},
  volume={23},
  number={11},
  pages={16992--17006},
}

@ARTICLE{DiRenzoRIS,
  author={Di Renzo, Marco and Zappone, Alessio and Debbah, Merouane and Alouini, Mohamed-Slim and Yuen, Chau and de Rosny, Julien and Tretyakov, Sergei},
  journal={{IEEE} J. Sel. Areas Commun.}, 
  title={{Smart Radio Environments Empowered by Reconfigurable Intelligent Surfaces: How It Works, State of Research, and The Road Ahead}}, 
  year={2020},
  volume={38},
  number={11},
  pages={2450--2525},
}

@INPROCEEDINGS{AnSIMMuBF,
  author={An, Jiancheng and Di Renzo, Marco and Debbah, Mérouane and Yuen, Chau},
  booktitle={ICC 2023}, 
  title={{Stacked Intelligent Metasurfaces for Multiuser Beamforming in the Wave Domain}}, 
  year={2023},
  volume={},
  number={},
  pages={2834--2839},
}

@ARTICLE{AnTransceiverDesign,
  author={An, Jiancheng and Yuen, Chau and Xu, Chao and Li, Hongbin and Ng, Derrick Wing Kwan and Di Renzo, Marco and Debbah, Mérouane and Hanzo, Lajos},
  journal={{IEEE} Wireless Commun.}, 
  title={{Stacked Intelligent Metasurface-Aided MIMO Transceiver Design}}, 
  year={2024},
  volume={31},
  number={4},
  pages={123--131},
}

@ARTICLE{DiRenzoIFFT,
  author={An, Jiancheng and Yuen, Chau and Guan, Yong Liang and Renzo, Marco Di and Debbah, Mérouane and Poor, H. Vincent and Hanzo, Lajos},
  journal={{IEEE} J. Sel. Areas Commun.}, 
  title={{Two-Dimensional Direction-of-Arrival Estimation Using Stacked Intelligent Metasurfaces}}, 
  year={2024},
  volume={42},
  number={10},
  pages={2786--2802},
}

@INPROCEEDINGS{TorcolacciEMProcessing,
  author={Fabiani, M. and Torcolacci, G. and Dardari, D.},
  booktitle={ASILOMAR 2025}, 
  title={Nonlinear EM-based Signal Processing}, 
  year={2025},
  volume={},
  number={},
  pages={1--6},
}

@ARTICLE{RiceanFading,
  author={Papazafeiropoulos, Anastasios and Kourtessis, Pandelis and Kaklamani, Dimitra I. and Venieris, Iakovos S.},
  journal={{IEEE} Wireless Commun. Lett.}, 
  title={Channel Estimation for Stacked Intelligent Metasurfaces in Rician Fading Channels}, 
  year={2025},
  volume={14},
  number={5},
  pages={1411--1415},
}

@ARTICLE{ChEstWireless,
  author={Yao, Xianghao and An, Jiancheng and Gan, Lu and Di Renzo, Marco and Yuen, Chau},
  journal={{IEEE} Wireless Commun. Lett.}, 
  title={Channel Estimation for Stacked Intelligent Metasurface-Assisted Wireless Networks}, 
  year={2024},
  volume={13},
  number={5},
  pages={1349--1353},
}

@INPROCEEDINGS{MisoChEst,
  author={Nadeem, Qurrat-Ul-Ain and An, Jiancheng and Chaaban, Anas},
  booktitle={WCNC 2024}, 
  title={{Hybrid Digital-Wave Domain Channel Estimator for Stacked Intelligent Metasurface Enabled Multi-User MISO Systems}}, 
  year={2024},
  volume={},
  number={},
  pages={1--6},
}

@ARTICLE{TensorChEst,
  author={Ginige, Nipuni and Sena, Arthur Sousa de and Mahmood, Nurul Huda and Renzo, Marco Di and Rajatheva, Nandana and Latva-Aho, Matti},
  journal={{IEEE} Commun. Lett.}, 
  title={{Nested Tensor-Based Channel Estimation for Stacked Intelligent Metasurface-Assisted Wireless Networks}}, 
  year={2025},
  volume={29},
  number={11},
  pages={2731--2735},
}

@ARTICLE{ChEstDL1,
  author={Lawal, Abdulmajid and Zerguine, Azzedine and Nasir, Ali A. and Abed-Meraim, Karim},
  journal={{IEEE} Commun. Lett.}, 
  title={{Channel Estimation for Stacked Intelligent Metasurface-Aided Network Using Deep Learning}}, 
  year={2025},
  volume={29},
  number={11},
  pages={2641--2645},
}

@ARTICLE{ChEstDL2,
  author={Dong, Xin and Chen, Chen and Yu, Gang and Zhou, Lingyou and Yuan, Chenyang and Zhang, Jie},
  journal={{IEEE} Trans. Veh. Technol.}, 
  title={{Deep Learning-Based Channel Estimation for Stacked Intelligent Metasurface-Enhanced Multi-User Communications}}, 
  year={2026},
  volume={},
  number={},
  pages={1--6},
}

@ARTICLE{SIMISAC1,
  author={Niu, Haoxian and An, Jiancheng and Papazafeiropoulos, Anastasios and Gan, Lu and Chatzinotas, Symeon and Debbah, Mérouane},
  journal={{IEEE} Wireless Commun. Lett.}, 
  title={{Stacked Intelligent Metasurfaces for Integrated Sensing and Communications}}, 
  year={2024},
  volume={13},
  number={10},
  pages={2807--2811},
}

@ARTICLE{SIMISAC2,
  author={Jiang, Chengjun and Yuan, Hao and Zhang, Chensi and An, Jiancheng and Huang, Chongwen and Yuen, Chau},
  journal={{IEEE} Wireless Commun. Lett}, 
  title={{Stacked Intelligent Metasurface-Enabled Satellite Integrated Sensing and Communications Systems}}, 
  year={2026},
  volume={15},
  number={},
  pages={930--934},
}

@INPROCEEDINGS{AbrardoEusipco,
  author={Abrardo, Andrea and Bartoli, Giulio and Toccafondi, Alberto and Di Renzo, Marco},
  booktitle={EUSIPCO 2025}, 
  title={{Leveraging Stacked Intelligent Surfaces for Near-Field Localization by Using a Multi-Port Network Model}}, 
  year={2025},
  volume={},
  number={},
  pages={1193--1197},
}

\end{document}